\documentclass[aps,showpacs,prl,twocolumn,longbibliography]{revtex4-2}

\usepackage[pdftex]{graphicx}
\usepackage{amssymb}
\usepackage{amsmath}
\usepackage{xspace}
\usepackage{color}
\usepackage{hyperref}

\usepackage[resetlabels]{multibib}
\newcites{SM}{Refs in Supplemental Material}

\newcommand{\tH}{t_{\rm h}}
\newcommand{\Lw}{L_{\scalebox{0.6}{\rm W}}}
\newcommand{\xM}{x_{\scalebox{0.6}{\rm M}}}

\begin{document} 

\title{Nonequilibrium dynamics in pumped Mott insulators}

\author{Satoshi Ejima}
\author{Florian Lange}
\author{Holger Fehske}
\affiliation{Institut f\"ur Physik, Universit\"at Greifswald, 17489 Greifswald, Germany}

\date{\today}

\begin{abstract}
We use time-evolution techniques for (infinite) matrix-product-states to calculate, directly in the thermodynamic limit, the time-dependent photoemission spectra and dynamic structure factors of the half-filled Hubbard chain after pulse irradiation. These quantities exhibit clear signatures of the photoinduced phase transition from insulator to metal that occurs because of the formation of so-called $\eta$ pairs.
In addition, the spin dynamic structure factor loses spectral weight in the whole momentum space, reflecting the suppression of antiferromagnetic correlations due to the buildup of $\eta$-pairing states.
The numerical method demonstrated in this work can be readily applied to other one-dimensional models driven out of equilibrium by optical pumping.
\end{abstract}

\maketitle

\textit{Introduction}--
The study of systems under optical excitation receives tremendous attention because of both the recent rapid developments of ultrafast pump lasers and the discovery of striking phenomena not observable in equilibrium, such as photoinduced superconducting-like states in high-$T_{\rm c}$ cuprates~\cite{Fausti2011,Hu2014,PhysRevB.89.184516} and the alkali-doped fulleride K$_3$C$_{60}$~\cite{Mitrano16,Budden21}, charge-density-waves in the rare-earth tritelluride LaTe$_3$~\cite{Kogar20}, or the insulator-to-metal transition in the excitonic-insulator candidate, Ta$_2$NiSe$_5$~\cite{PhysRevLett.119.086401,Okazaki2018,PhysRevB.101.235148}.
Pump-probe experiments allow to explore various physical processes at different time scales. In the pre-pump region the system can be still described by linear-response theory. On the other hand, the recovery process of the system in the long-time limit after a pump provides 
valuable information about lifetimes and interaction mechanisms of the system. Here, we are interested in the rich nonequilibrium physics during or shortly after the pump. 
Time-dependent spectroscopic measurements, such as a time- and angle-resolved photoemission spectroscopy (TARPES)~\cite{Zhou2018,NatRevPhys1-609}, in principle enable direct comparison with theory in this time domain.  
It is, however, challenging to tackle these problems by numerical techniques, especially in systems with emergent photoinduced phase transitions.  

Various numerical techniques have been applied to optically excited systems to study nonequilibrium dynamics, e.g., the exact-diagonalization technique~\cite{Park1986,PhysRevB.96.235142,PhysRevB.96.235142,Okamoto19}, extensions of dynamical mean-field theory (DMFT)~\cite{RevModPhys.86.779} and dynamical cluster approaches~\cite{PhysRevB.101.085127}.
While ED simulations can access long times but are limited to small clusters (so far with less than 14 sites in fermionic Hubbard systems), the time-dependent DMFT is constructed directly in the thermodynamic limit but restricted by the local approximation for the self-energy, which may be inappropriate in low-dimensional systems with inherent nonlocal spatial correlations.
In one dimension, methods based on matrix-product-states (MPSs), such as the density-matrix renormalization group (DMRG)~\cite{White92,Sch11} and its time-dependent version~\cite{PhysRevLett.93.076401,Daley2004}, enable us to simulate both static and dynamic properties with high accuracy even for large systems. 
Results for nonequilibrium dynamics are still rare, however. 
The time evolution of the spectral functions after a quantum quench has been computed in the extended Hubbard chain at half filling for finite systems~\cite{PhysRevB.100.195124,PhysRevB.101.180507,PhysRevB.102.235141,PhysRevB.104.085122,Meyer21}. Very recently, photodoped Mott insulators have been studied by an infinite time-evolving block decimation (iTEBD) approach~\cite{TEBD,iTEBD} with the help of a generalized Gibbs ensemble to obtain an appropriate effective model description at equilibrium~\cite{Murakami22}. 

In this paper, we propose a direct numerical scheme for the computation of nonequilibrium dynamic response functions, which can be used for general (quasi-)one-dimensional (1D) systems. We apply this technique to the optically excited half-filled Hubbard chain, in which a photoinduced $\eta$-pairing state appears~\cite{Kaneko19}. In the time-resolved photoemission spectra (PES), an extra band becomes visible above the Fermi energy after pulse irradiation, indicating the insulator-to-metal phase transition.
Further evidence for this transition can be found in the charge dynamic structure factor (DSF), where a gapless band appears.

\textit{Nonequilibrium dynamics}---
To explore the system's dynamics at nonequilibrium, time-dependent spectral functions of the form~\cite{PhysRevLett.102.136401,PhysRevB.96.235142,Okamoto19} 
\begin{align}
I(k,\omega;t) = \sum_{r}e^{-\mathrm{i}kr} \int_{-\infty}^{\infty}\int_{-\infty}^{\infty} d\tau_1 d\tau_2 f(\tau_1,\tau_2;\omega)
\nonumber \\
\times C(r,\tau_1,\tau_2; t) 
\label{noneq-dyn}
\end{align}
are of interest. Here, $C(r,\tau_1,\tau_2; t)=\langle\phi(t)|\hat{O}_{j+r}^\dagger(\tau_1;t)\hat{O}_{j}(\tau_2;t)|\phi(t)\rangle$ is the nonequilibrium two-point correlator at times $\tau_1$ and $\tau_2$ defined relative to time $t$ for a local operator $\hat{O}_j$. The prefactor is given by 
$f(\tau_1,\tau_2;\omega)=e^{{\mathrm i}\omega(\tau_1-\tau_2)}g(\tau_1)g(\tau_2)$ with 
$g(\tau)=\exp[-\tau^2/2\sigma_{\rm pr}^2]/\sqrt{2\pi}\sigma_{\rm pr}$ describing the shape of a probe pulse, e.g., in a time-dependent photoemission spectroscopy experiment. 

We calculate the correlator $C(r,\tau_1,\tau_2; t)$ numerically by simulating states 
$\hat{O}_j(\tau;t)|\phi(t)\rangle=\hat{U}^\dagger(t+\tau,t) \hat{O}_j \hat{U}(t+\tau,t)|\phi(t)\rangle\equiv |\bar{\phi}(\tau;t)\rangle$ within an (i)MPS representation. 
Here, $\hat{U}(t+\tau,t)\equiv {\cal T}\exp[-\mathrm{i}\int_t^{t+\tau} dt^\prime \hat{H}(t^\prime)]$ is the unitary time-evolution operator of the system with the (reverse) time-ordering operator ${\cal T}$ for $\tau>0$ ($\tau<0$).
To this end, we first prepare the iMPS approximation $|\psi_0\rangle$ of the ground state of the unperturbed Hamiltonian $\hat{H}$ by infinite DMRG~\cite{Mc08}.
We then utilize the iTEBD technique to carry out a time evolution according to the time-dependent Hamiltonian  $\hat{H}(t)$, which includes an electric field $A(t)$, in order to obtain an iMPS representation of the photoinduced state $|\psi(t)\rangle$. 
Using the iMPS description reduces the numerical cost, since the size of the unit cell $N_{\rm uc}$ is usually much smaller than the system sizes needed in simulations with open boundary conditions.
Applying the operator $\hat{O}_j$ lifts the translation invariance of the state, however, so we need to switch to an MPS representation with \textit{infinite boundary conditions} (IBCs) for the rest of the simulation. 
More specifically, `Method I' described in Ref.~\cite{Zauner2015} is necessary because of the photoexcited state is not an eigenstate of the Hamiltonian that determines the time evolution. 
Having calculated states $|\bar{\phi}(\tau;t)\rangle$ for times $\tau_1$ and $\tau_2$, the two-point correlators $C(r,\tau_1,\tau_2;t)$ in Eq.~\eqref{noneq-dyn} can be evaluated by shifting $|\bar{\phi}(\tau_1;t)\rangle$ and $|\bar{\phi}(\tau_2;t)\rangle$ relative to each other~\cite{Spin1FinT,Ejima21}. In this way, the number of MPSs $|\bar{\phi}(\tau;t)\rangle$ required for some fixed time $\tau$ is drastically reduced. Namely, we only need a number of MPSs proportional to $N_{\rm uc}$, instead of $L$ as in a similar approach with open boundary conditions. 
Further technical details are given in the Supplementary Materials~\cite{SM}.

\textit{Model}---
Our target system is the 1D half-filled Hubbard model with nearest-neighbor hopping $\tH$ and on-site Coulomb repulsion $U>0$,
\begin{align}
 \hat{H}=&
 -\tH \sum_{j,\sigma}
  \big(
   \hat{c}_{j,\sigma}^\dagger \hat{c}_{j+1,\sigma}^{\phantom{\dagger}}
   +\text{H.c.}
  \big)
  \nonumber \\
 &
  +U\sum_{j}\left(
     \hat{n}_{j,\uparrow}-1/2
    \right)
    \left(
    \hat{n}_{j,\downarrow}-1/2
   \right)\,,
  \label{hubbard}
\end{align}
where $\hat{c}_{j,\sigma}^{\dagger}$ ($\hat{c}_{j,\sigma}^{\phantom{\dagger}}$)
creates (annihilates) a fermion with spin projection $\sigma$ ($=\uparrow,\downarrow$) at lattice site $j$, and  $\hat{n}_{j,\sigma}=\hat{c}_{j,\sigma}^{\dagger} \hat{c}_{j,\sigma}^{\phantom{\dagger}}$.
The ground state for $U>0$ is a Mott insulator with a finite charge gap $\Delta$. 
As Yang demonstrated in a seminal paper~\cite{Yang89}, exact eigenstates of the Hubbard model can be constructed 
by means of the so-called $\eta$-operators 
\begin{align}
 \hat{\eta}^+ &= \sum_j(-1)^j \hat{\Delta}_j^\dagger
 \equiv \sum_j \hat{\eta}_j^+\, ,
 \ \  \ 
 \hat{\eta}^- = (\hat{\eta}^+)^\dagger\, ,
 \label{etapm}
\\
 \hat{\eta}^z &= \frac{1}{2}\sum_{j}  (\hat{n}_{j,\uparrow}+\hat{n}_{j,\downarrow}-1)
 \equiv \sum_j \hat{\eta}_j^z \, ,
 \label{etaz}
\end{align}
which obey the SU(2) commutation relations. $\hat{\Delta}_j^\dagger=\hat{c}_{j,\downarrow}^\dagger\hat{c}_{j,\uparrow}^\dagger$ is the singlet pair creation operator. 

The Hubbard Hamiltonian~\eqref{hubbard} commutes with the operator $\hat{\eta}^2=\tfrac{1}{2}(\hat{\eta}^+\hat{\eta}^-+\hat{\eta}^-\hat{\eta}^+) +  (\hat{\eta}^z)^2$, so that $\langle \eta^2 \rangle$ is a conserved quantity in the absence of perturbations. Eigenstates with a finite value of $\langle \hat{\eta}^2 \rangle$ have long-ranged pairing correlations $\langle\hat{\eta}^+_j\hat{\eta}^-_{\ell}\rangle$~\cite{Yang89}. 
While these $\eta$-pairing states cannot be ground states, it has been recently recognized that pulse irradiation can induce $\eta$-pairing in Mott insulators~\cite{Kaneko19}. Here, we study such photoinduced $\eta$-pairing states with the time-evolution technique outlined above. 

\begin{figure}[tb]
  \includegraphics[width=0.99\linewidth]{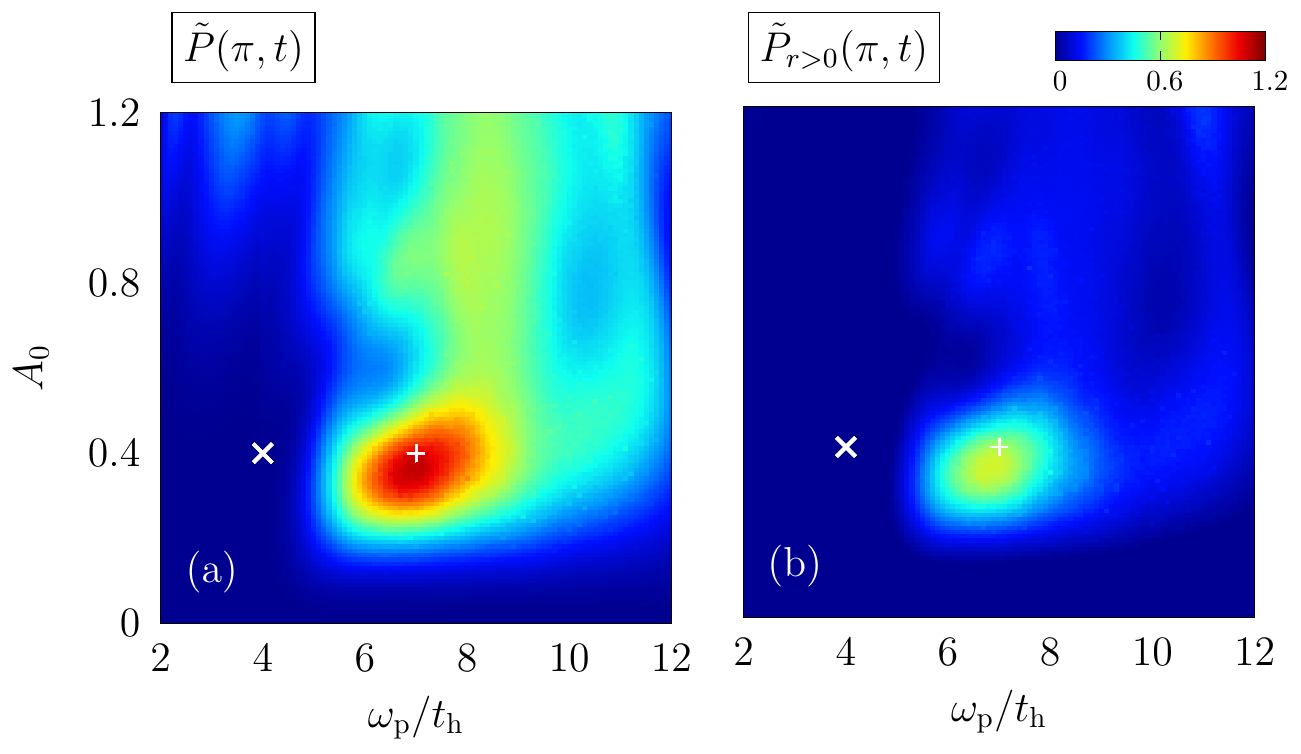}
  \caption{
  Contour plots of $\tilde{P}(\pi,t=15\tH^{-1})$ (a) and $\tilde{P}_{r>0}(\pi,t=15\tH^{-1})$ (b) in the $\omega_{\rm p}$-$A_0$ plane for an infinite Hubbard chain  with $U/\tH=8$ at half filling. The pump pulse has width $\sigma_{\rm p}=2$ and is centered at time $t_0\cdot \tH=10$. }
  \label{contour-plots}
\end{figure}

We introduce the external time-dependent electric field $A(t)$  via a Peierls phase~\cite{Peierls1933} as 
$\tH \hat{c}_{j,\sigma}^\dagger \hat{c}_{j+1,\sigma}^{\phantom{\dagger}}\to \tH e^{\mathrm{i}A(t)} \hat{c}_{j,\sigma}^\dagger \hat{c}_{j+1,\sigma}^{\phantom{\dagger}}$, 
where 
\begin{eqnarray}
 A(t)=A_0 e^{-(t-t_0)^2/(2\sigma_{\rm{p}}^2)}\cos\left[\omega_{\rm{p}}(t-t_0)\right] \, .
 \label{pump}
\end{eqnarray}
This describes a pump pulse with amplitude $A_0$, frequency $\omega_{\rm p}$ and
width  $\sigma_{\rm p}$, centered at time $t_0$ ($>0$). 
As a result, the Hamiltonian becomes time dependent, $\hat{H}\to\hat{H}(t)$, 
and the system being initially in the ground state is driven out of equilibrium: $|\psi(0)\rangle\to |\psi(t)\rangle$. 

The $\eta$-pairing state can be detected by evaluating the time evolution of the pair-correlation function
\begin{align}
P(r,t)=\frac{1}{L}\sum_j \langle \psi(t)|\hat{\Delta}^\dagger_{j+r}\hat{\Delta}_j^{\phantom{\dagger}}+{\rm H.c.}) |\psi(t)\rangle\,
\end{align}
and its Fourier transform $\tilde{P}(q,t)=\sum_r e^{\mathrm{i}q r} P(r,t)$.
As demonstrated in Refs.~\cite{Kaneko19,SCES19} for small clusters, and in Ref.~\cite{Ejima20} in the thermodynamic limit, $\tilde{P}(\pi,t)$ is enhanced after pulse irradiation, which implies the formation of an $\eta$-pairing state.
The optimal parameter set for inducing $\eta$-pairing 
thus can be determined examining the $A_0$- and $\omega_{\rm p}$-dependences of $\tilde{P}(\pi,t)$ with the iTEBD technique~\cite{Ejima20}. 

Figure~\ref{contour-plots}(a) shows the contour plot of $\tilde{P}(\pi,t)$ after pulse irradiation ($t\cdot \tH=15$). 
Obviously, there is a maximum around $A_0\approx0.4$ and $\omega_{\rm p}/\tH \approx 7.0$ marked by `+'. In the following, we analyze the nonequilibrium spectral functions for this parameter set. 
To demonstrate that the nonlocal part of the pairing correlations is dominant for these optimal parameters, we also extract the contour plot of the modified structure factor $\tilde{P}_{r>0}(q,t)=\sum_{r>0}e^{\mathrm{i}qr} P(r,t)$, in which the contribution of the double occupancy 
$n_{\rm d}(t)=(1/L)\sum_j \langle \psi(t)|\hat{n}_{j,\uparrow}\hat{n}_{j,\downarrow}|\psi(t)\rangle$ 
is excluded [Fig.~\ref{contour-plots}(b)]. Clearly, the spectral intensity of $\tilde{P}_{r>0}(\pi,t)$ is only appreciable in the vicinity of `+'. 
We also explore the nonequilibrium dynamics away from the optimal parameter set, at $A_0=0.4$ and $\omega_{\rm p}=4.0$ (denoted as `$\times$` in Fig.~\ref{contour-plots}), to elucidate the difference between $\eta$-pair-correlation dominant and non-dominant regions.

\textit{Photoemission spectra}---
\begin{figure}[t]
  \includegraphics[width=0.99\linewidth]{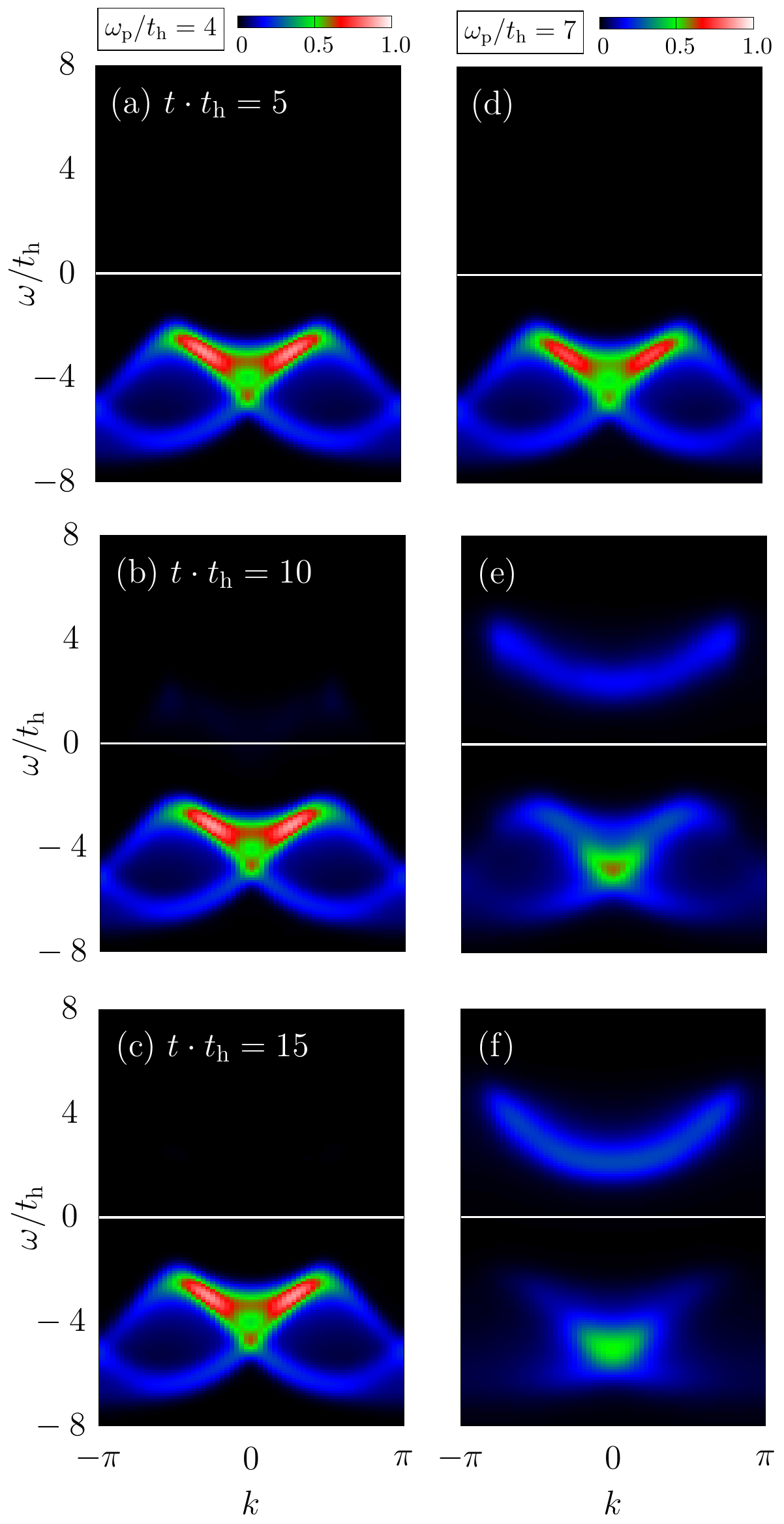}
  \caption{Snapshots of the photoemission spectra $A^-(k,\omega; t)$ for the $\eta$-pairing non-dominant (left panels with $\omega_{\rm p}/\tH=4$) and dominant (right panels with $\omega_{\rm p}/\tH=7$) states during the pump at $t\cdot\tH=5$ [(a) and (d)], $10$ [(b) and (e)], and $15$ [(c) and (f)]. The data are obtained by the (i)TEBD technique with IBCs at $U/\tH=8$, where the pump is parametrized by $A_0=4$ and $\sigma_{\rm p}=2$ at $t_0\cdot\tH=10$.
 }
  \label{Akwt}
\end{figure}
Since the 1D half-filled Hubbard model has an insulating ground state for any $U>0$, there exists a Mott gap ($\Delta\sim U$) between lower and upper Hubbard bands in the single-particle spectral function $A(k,\omega)=A^+(k, \omega)+A^-(k,\omega)$, where $A^-(k,\omega)$ and $A^+(k,\omega)$ denote the PES and inverse PES, respectively.  
In the superconducting $\eta$-pairing state after pulse irradiation, however, the Mott gap is melted, which should also be captured in the time-dependent single-particle spectral function $A(k,\omega; t)$. Because of particle-hole symmetry, $A^+(k,\omega; t)$ and $A^-(k,\omega; t)$ contain the same information. We focus on $A^-(k,\omega; t)$, which is obtained by setting $\hat{O}_j = \hat{c}_{j,\sigma}$ in Eq.~\eqref{noneq-dyn}.

Figure~\ref{Akwt} displays our (i)TEBD results for $A^-(k,\omega; t)$, using the pump parameter sets `$\times$' and `+' of Fig.~\ref{contour-plots}, which correspond to angular frequencies $\omega_{\rm p}/\tH = 4$ and $\omega_{\rm p}/\tH = 7$, respectively.
For $\omega_{\rm p}/\tH = 4$, the time-dependent spectral function is very similar to the equilibrium spectral function at $T=0$ (see, e.g., Ref~\cite{Ejima21}), i.e., it is only slightly changed by the pulse irradiation. 
In contrast, when the parameters optimized to induce $\eta$-pairing are used ($\omega_{\rm p}/\tH = 7$), an extra dispersion above the Fermi energy ($\omega>E_{\rm F}$) appears during the pump and persists afterwards [Fig.~\ref{Akwt}(e)-(f)].

\begin{figure}[tb]
  \includegraphics[width=0.99\linewidth]{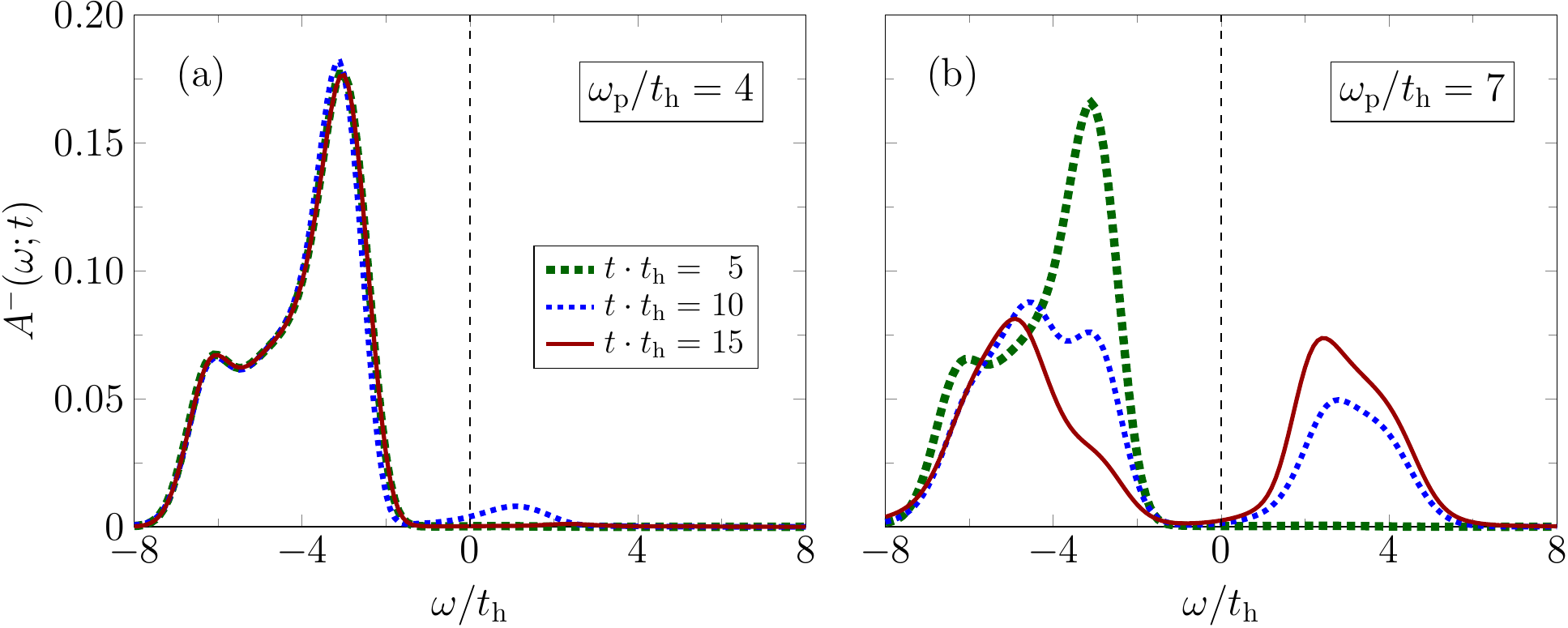}
  \caption{The transient integrated density of states $A^-(\omega; t)$ from Eq.~\eqref{EQ-DOS} for $U/\tH=8$ with $\omega_{\rm p}/\tH=4$ (a) and  $7$ (b). The pump is parametrized as in Fig.~\ref{Akwt}. 
  }
  \label{Awt}
\end{figure}

Evaluating the integrated density of states
\begin{align}
A^-(\omega; t)=\tfrac{1}{L}\sum_k A^-(k,\omega; t)\,,
\label{EQ-DOS}
\end{align}
we see more clearly how the spectral weight is shifted from $\omega<E_{\rm F}$ to $\omega>E_{\rm F}$ by the photoinduced $\eta$-pairing. Figure~\ref{Awt}(a) shows $A^-(\omega;t)$ for $\omega_{\rm p}/\tH=4$. Although a small shift of the spectral weight to $\omega>E_{\rm F}$ is observed at $t\approx t_0$, it becomes negligible after the pulse ($t\cdot\tH\gtrsim 15$). 
On the other hand, the spectral weight for $\omega>E_{\rm F}$ increases distinctly over time in the $\omega_{\rm p}/\tH=7$ case [Fig.~\ref{Awt}(b)], indicating a photoinduced phase transition from a Mott insulator to a metallic $\eta$-pairing state.
This photoinduced insulator-to-metal transition should be observed in TARPES, when the pure Hubbard model is realized experimentally, e.g. in optical lattices (although this would require further developments of ARPES techniques~\cite{Brown2019}).

\textit{Dynamic structure factor}---
We now analyze the spin and charge DSFs, $S_{\rm s}(q,\omega; t)$ and $S_{\rm c}(q,\omega; t)$, which are obtained by setting $\hat{O}_j=\hat{S}_j^z$ and $\hat{O}_j=\hat{n}_{j\uparrow}+\hat{n}_{j\downarrow}-1$ in Eq.~\eqref{noneq-dyn}, respectively. 
\begin{figure}[tb]
  \includegraphics[width=0.99\linewidth]{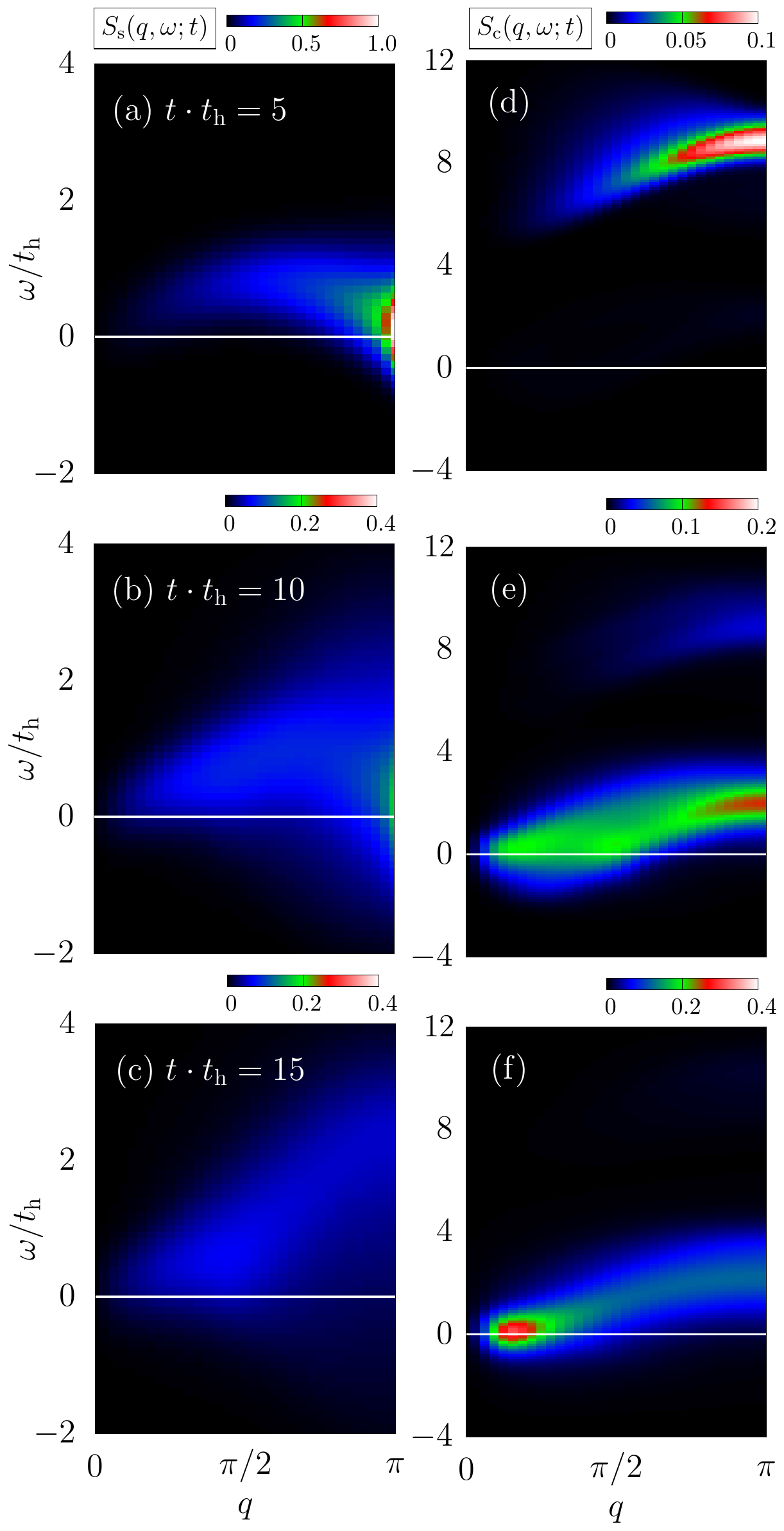}
  \caption{Snapshots of the dynamic structure factors for spin [$S_{\rm s}(q,\omega; t)$] and charge [$S_{\rm c}(q,\omega; t)$]
  during the pump at $t\cdot\tH=5$ [(a) and (d)], $10$ [(b) and (e)], and $15$ [(c) and (f)].
  The data are obtained by the time-evolution technique with IBCs, where the pump is parametrized by $A_0=0.4$ and $\omega_{\rm p}/\tH=7$ (`+' symbol in Fig.~\ref{contour-plots}).
  }
  \label{td-DSF}
\end{figure}

Let us first recall the character of the DSF at zero temperature. 
For any $U>0$, the spin DSF consists of a two-spinon continuum with an excitation gap that closes at momenta $q=0$ and $\pi$, while the charge DSF is gapful, reflecting the Mott gap for holon excitations (see also the numerical results with IBCs in Ref.~\cite{SM}). 

Figures~\ref{td-DSF}(a)-(c) give the nonequilibrium spin DSF $S_{\rm s}(q,\omega; t)$ in the half-filled Hubbard model with $U/\tH=8$, using the optimal parameter set (`+' from Fig.~\ref{contour-plots}). 
In Fig.~\ref{td-DSF}(a) $S_{\rm s}(q,\omega; t)$ is quite similar to $S_{\rm s}(q,\omega)$ at equilibrium, showing the gap closing at the momenta $q=0$ and $\pi$. During and after pulse irradiation [Figs.~\ref{td-DSF}(b) and (c)] the spectral weight is reduced in the whole momentum space, reflecting the suppression of antiferromagnetic correlations in the photoinduced $\eta$-pairing state~\cite{Kaneko19}.

\begin{figure}[t]
 \includegraphics[width=0.8\linewidth]{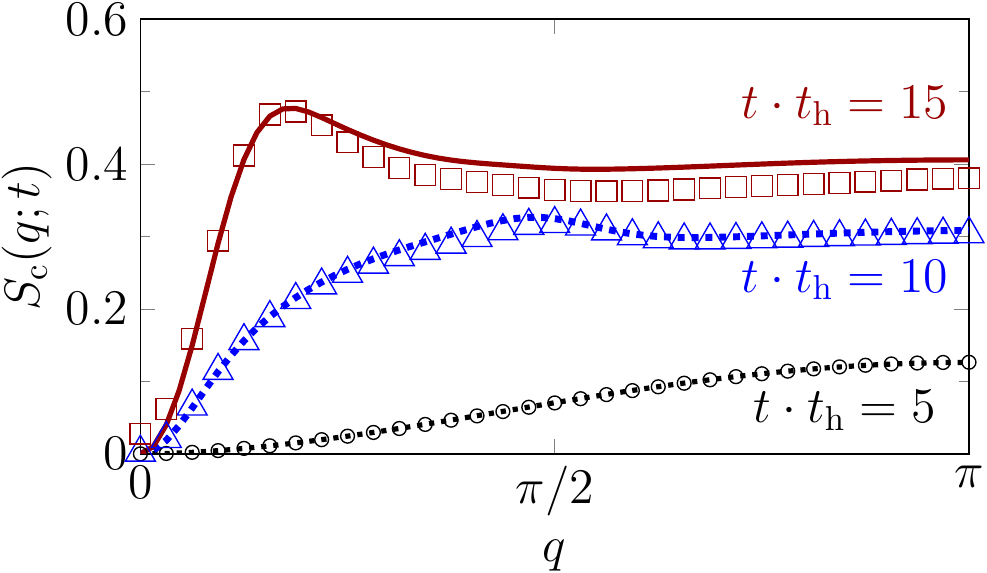}
 \caption{The time-dependent charge structure factor $S_{\rm c}(q;t)$ for $\omega_{\rm p}/\tH=7$ obtained from the energy-integrated $S_{\rm c}(q,\omega;t)$ (symbols)
 and the Fourier transform of 
$\langle \psi(t)|\hat{\eta}_{j+r}^z\hat{\eta}_{j}^z|\psi(t)\rangle$ [Eq.~\eqref{Scq}]
 (lines), where the pump is parametrized as in Fig.~\ref{td-DSF}.
}
 \label{Sc-qt}
\end{figure}

More drastic changes should be expected in the charge DSF, since $S_{\rm c}(q,\omega)$ can be written with the $\eta$-operators in Eqs.~\eqref{etapm} and \eqref{etaz} as discussed in Refs.~\cite{SM,PhysRevB.85.165132}.
 This is in accordance with the numerical results shown in Figs.~\ref{td-DSF}(d)-(e).
Before the pump pulse [Fig.~\ref{td-DSF}(d)], $S_{\rm c}(q,\omega; t)$ is nearly indistinguishable from the equilibrium DSF $S_{\rm c}(q,\omega)$ at $T=0$, with the Mott gap ($\Delta_{\rm c}\approx 4.68$ for $U/\tH=8$) visible at $q=0$. 
During the pump ($t=t_0$) an extra band appears and closes the gap [Fig.~\ref{td-DSF}(e)].
Most significantly, the spectral weight now concentrates at $q\approx0.55$ and $\omega\approx 0$, which is confirmed more clearly in the energy-integrated charge structure factor $S_{\rm c}(q;t) = \int_{-\infty}^\infty d\omega S_{\rm c}(q,\omega;t)$, as shown in Fig.\ref{Sc-qt}. Note that $S_{\rm c}(q;t)$ can also be obtained from the Fourier transform of the two-point correlation functions 
$\langle \psi(t) | \hat{\eta}_{j+r}^z \hat{\eta}_j^z|\psi(t) \rangle$ 
  using iTEBD in the iMPS representation:
  \begin{align}
  S_{\rm c}(q;t)&= 4\int_{-\infty}^\infty d\tau 
 \frac{e^{-\tau^2 / \sigma_{\rm pr}^2}}{ \sigma_{\rm pr}^2} 
 \sum_r e^{-\mathrm{i}qr} \langle\psi(t)|\hat{\eta}_{j+r}^z\hat{\eta}_{j}^z|\psi(t)\rangle
 \nonumber \\
 &= \frac{4\sqrt{\pi}}{\sigma_{\rm pr}}
 \sum_r e^{-\mathrm{i}qr} \langle\psi(t)|\hat{\eta}_{j+r}^z\hat{\eta}_{j}^z|\psi(t)\rangle\,.
 \label{Scq}
\end{align}
This peak structure might be taken as an indication for superfluidity of the $\eta$-pairing state. 
It would be desirable to extract an order parameter with the help of field theoretical analysis analogous to the Tomonaga-Luttinger liquid parameter in the 1D Bose-Hubbard model, which characterizes the superfluid phase, see, e.g., Ref.~\cite{Ejima2011}.
Note that in Ref.~\cite{PhysRevB.101.180507}, a related two-particle spectral function [$\hat{O}_j=\hat{\Delta}_j$] was studied to show a nonequilibrium transition to a superconducting phase.

\textit{Conclusions}---
We have demonstrated how the spectral functions of (quasi-)one-dimensional systems at nonequilibrium can be simulated directly in the thermodynamic limit by using the time-dependent density-matrix renormalization group technique with infinite boundary conditions. 
We have applied this technique to the optically excited Hubbard chain at half filling and observed that so-called $\eta$-pairing states appear after pulse irradiation. 
Tuning the pump pulse to maximize $\eta$-pairing, we have found clear evidence for a photoinduced metal-insulator transition in both the time-dependent photoemission spectra and dynamic structure factors.

Some care should be taken when using the spectral function $I(k,\omega;t)$ in Eq.~\eqref{noneq-dyn} to interpret TARPES experiments, since its derivation relies on the assumption that the pump and probe pulses do not overlap. For overlapping pulses, gauge invariance may be violated~\cite{Freericks2015}. 
 Importantly, however, this does not affect the results for long times ($t\cdot\tH = 15$), which are the most relevant to our conclusions. Moreover, there is no problem related to gauge invariance for the integrated density of states shown in Fig.~\ref{Awt}, which is thus also valid.

The proposed numerical scheme opens a new venue for exploring nonequilibrium dynamics with high resolution and controllable accuracy. It would be of specific interest to try to numerically reproduce the experimental results of time-dependent ARPES of Ta$_2$NiSe$_5$~\cite{PhysRevLett.119.086401,Okazaki2018,PhysRevB.101.235148}, which is a strong candidate for excitonic insulators and exhibits the characteristic flat-band behavior in ARPES experiments~\cite{PhysRevLett.103.026402,Wakisaka2012} within a narrow region in momentum space~\cite{Ejima21}.

\vspace*{1cm}
\textit{Acknowledgments} ---
The iTEBD simulations were performed using the ITensor library~\cite{ITensor}.
S.E. and F.L. were supported by Deutsche Forschungsgemeinschaft through project 
EJ 7/2-1 and FE 398/8-1, respectively.

%


\clearpage


\pagebreak
\widetext
\begin{center}
 \textbf{\large Supplemental Materials: Nonequilibrium dynamics in pumped Mott insulators}
\end{center}

\vspace*{5pt}

\begin{center}
 Satoshi Ejima, Florian Lange, and Holger Fehske\\[5pt]
 \textit{Institut f\"ur Physik, Universit\"at Greifswald, 17489 Greifswald, Germany}
 \end{center}

\setcounter{equation}{0}
\setcounter{figure}{0}
\setcounter{table}{0}
\setcounter{page}{1}
\makeatletter
\renewcommand{\theequation}{S\arabic{equation}}
\renewcommand{\thefigure}{S\arabic{figure}}
\renewcommand{\bibnumfmt}[1]{[S#1]}
\renewcommand{\citenumfont}[1]{S#1}

\section{Numerical technique}

\begin{figure}[tbh]
  \includegraphics[width=0.99\linewidth]{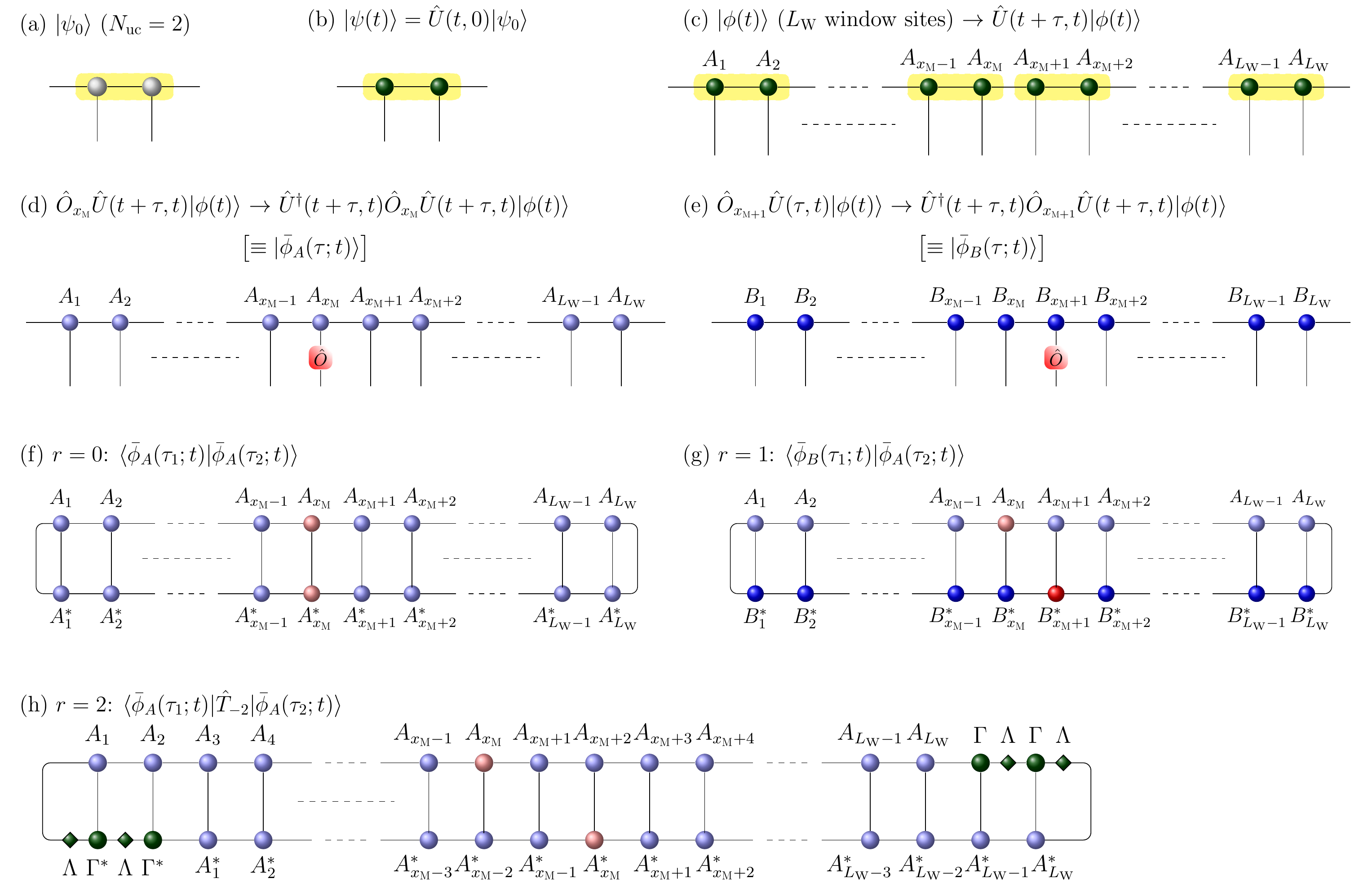}
  \caption{
  Graphical representation of the (i)MPSs used in the calculation of time-dependent dynamical correlation functions 
  $C(r,\tau_1,\tau_2;t)$. 
  (a) The iMPS for the ground state $|\psi_0\rangle$ is obtained by iDMRG. (b) The iMPS after time evolution $|\psi(t)\rangle$ is computed by iTEBD. (c) An MPS with $L_{\rm W}$ window sites $|\phi(t)\rangle$ can be constructed by the iMPS unit cell in (b). 
  After forward time evolution we apply the target operators $\hat{O}_{x_{\rm M}}$ and $\hat{O}_{x_{\rm M}+1}$ as in panel (d) and (e), respectively, and evolve the resulting MPSs backwards in time to prepare the MPSs $|\bar{\phi}_{A/B}\rangle$. The correlation functions can then be calculated for arbitrary distance, see e.g., panel (f) for $r=0$,  (g) for $r=1$, and especially panel (h) for $r=2$ by shifting the states relative to each other, i.e., by inserting two-site translation operator $\hat{T}_{-2}$. For further explanation see text.
    }
 \label{ibc}
\end{figure}

In this section, we explain how to extract the time-dependent correlation functions of the photoinduced states 
$C(r,\tau_1,\tau_2; t)=\langle\phi(t)|\hat{O}_{j+r}^\dagger(\tau_1;t)\hat{O}_{j}(\tau_2;t)|\phi(t)\rangle$ 
in Eq.~\eqref{noneq-dyn} using time-evolution techniques for (infinite) matrix-product states [(i)MPSs]. 
The whole scheme of our numerical simulations is as follows:

\vspace*{0.5cm}

\begin{enumerate}
	\item[0)] Calculate an iMPS representation of the ground state $|\psi_0\rangle$ using the infinite density-matrix renormalization group (iDMRG) [Fig.~\ref{ibc}(a)]. Here, we consider a two-site unit cell iMPS in the canonical form with tensors $\Lambda^{[n]}$ and $\Gamma^{[n]}$ ($n\in \{0,1\}$). Then evolve $|\psi_0\rangle$ in time by the infinite time-evolving block decimation (iTEBD) algorithm as
      $|\psi(t)\rangle=\hat{U}(t,0)|\psi_0\rangle$ [Fig.~\ref{ibc}(b)]. 
 \item[1)] Construct an MPS with infinite boundary conditions (IBCs) with an appropriate window size $\Lw$
       by repeating the unit cell of the iMPS $|\psi(t)\rangle$, see Fig.~\ref{ibc}(c). This state can be written as
      \[
       |\phi(t)\rangle=\sum_{\boldsymbol \sigma}\cdots\Lambda^{[1]}\Gamma^{[0]{\sigma_0}}
       A^{[1]\sigma_1}\cdots A^{[\Lw]\sigma_{\Lw}} \Gamma^{[1]\sigma_1}
      \Lambda^{[1]}\cdots|{\boldsymbol \sigma}\rangle\,,
      \]
       where $\sigma_j$ denotes the basis states of the local Hilbert space at site $j$. 
 \item[2)] Apply the target operators $\hat{O}_{\xM}$ and $\hat{O}_{\xM+1}$ to the MPS $\hat{U}(t+\tau,t)|\phi(t)\rangle$ [Figs.~\ref{ibc}(d) and (e)]
      and evolve the resulting states backwards in time as
      \[
       |\bar{\phi}_{A/B}(\tau;t)\rangle=\hat{U}^\dagger(t+\tau,t)\hat{O}_{\xM/\xM+1}\hat{U}(t+\tau,t)|\phi(t)\rangle
      \]
      with $\xM\equiv \Lw/2$. 
      For a nonequilibrium state $|\phi(t)\rangle$, the time evolution changes the state not only in the vicinity of the operator $\hat{O}_{\xM/\xM+1}$. The iMPS $|\psi(t)\rangle$ should therefore also be evolved in time, following `Method I' of Ref.~\citeSM{Zauner2015SM}, to update the tensors outside of the $\Lw$-site window. 
      Store the states $|\bar{\phi}_{A/B}(\tau ;t)\rangle$ for evenly spaced times $\tau \in [-T,T]$ to evaluate the two-point correlators later.      
 \item[3)] Evaluate the two-point correlation functions
      \[
       \langle \phi(t) |\hat{O}_j^\dagger(\tau_1;t)\hat{O}_\ell^{\phantom{\dagger}}(\tau_2;t)|\phi(t)\rangle
      \]
      for $-T \leq \tau_1,\tau_2 \leq T$ by shifting $|\bar{\phi}_{A/B}(\tau_1 ; t)\rangle$ and $|\bar{\phi}_{A/B}(\tau_2 ; t)\rangle$ relative to each other [see Figs.~\ref{ibc}(f)-(h) and App. A of Ref.~\citeSM{Ejima21SM}]. This is possible, because the state $|\phi (t) \rangle$ is symmetric under translation by one unit cell. Since our iMPS has a two-site unit cell, we need two sets of states, $|\bar{\phi}_{A}(\tau ; t)\rangle$ and $|\bar{\phi}_{B}(\tau ; t)\rangle$, to obtain the correlation function for all distances $r$. 
\end{enumerate}

\vspace*{0.5cm}

Note that in the case of the single-particle spectral functions $A(k,\omega;t)$ the Jordan-Wigner strings
$\hat{F}_j=\exp(\mathrm{i}\pi\hat{n}_j)$
with $\hat{n}_j=\hat{n}_{j,\uparrow}+\hat{n}_{j,\downarrow}$,
which appear in the mapping of spinful fermionic operators
into spinful bosonic operators, need to be taken into account
when preparing the MPSs for the time evolution (see Ref.~\citeSM{Ejima21SM} for details).

In our (i)TEBD simulations, we used a second-order Suzuki-Trotter decomposition with time step $0.1\tH^{-1}$  ($0.01\tH^{-1}$). The maximum MPS bond dimension was 1600. Furthermore, we chose $T=5\tH^{-1}$ as the time cutoff in the calculation of the time-dependent correlation functions, i.e., the integration in Eq.~\eqref{noneq-dyn} was done over $-T \leq \tau_1,\tau_2 \leq T$.

\section{Dynamic structure factor at zero temperature without pumping}
\label{DSF-T0}

\begin{figure}[tbh]
  \includegraphics[width=0.5\linewidth]{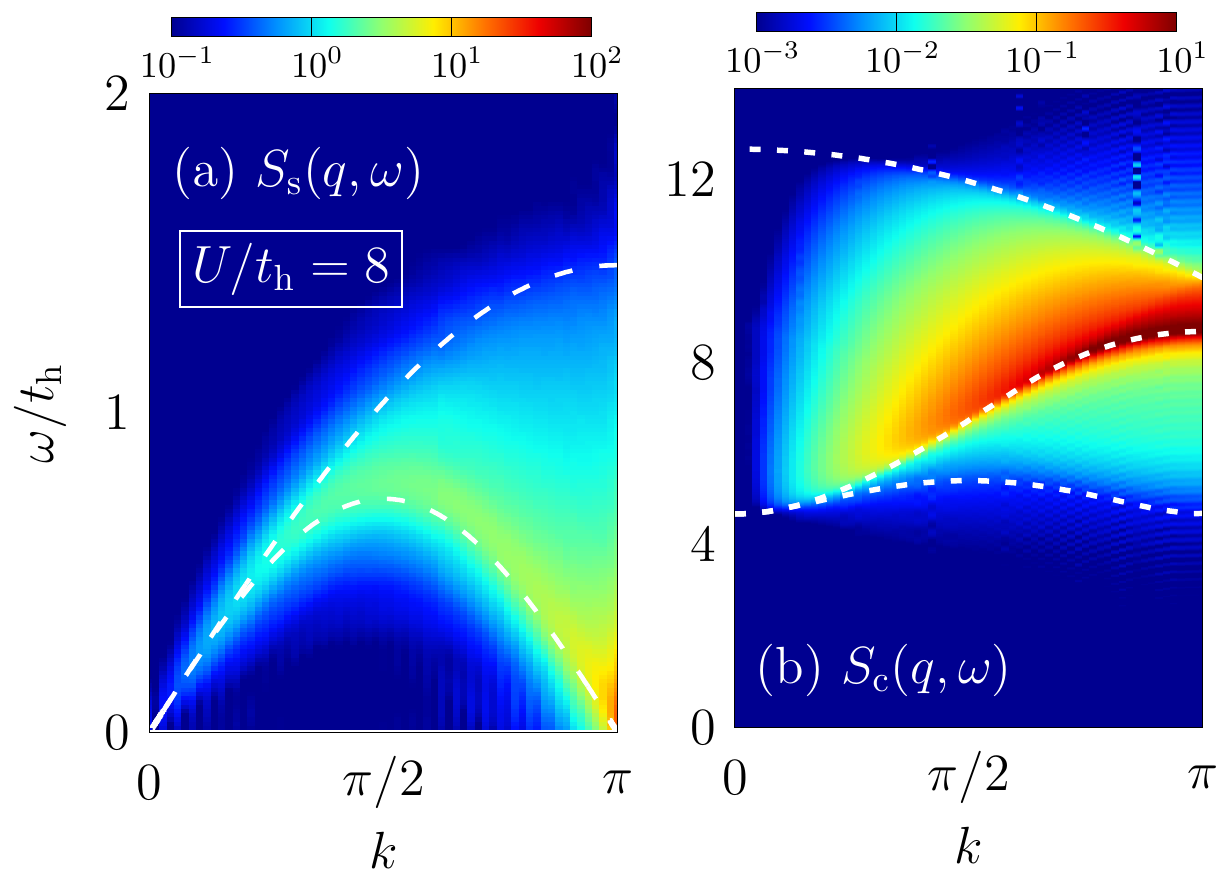}
  \caption{Spin (a) and charge (b) structure factors in the half-filled Hubbard model~\eqref{hubbard} with $U/\tH=8$ at zero temperature. 
  Dashed lines in panel (a) are upper and lower boundaries of the spinon-continuum, $\omega_{\rm up/low}(q)$. 
  Dotted lines in panel (b) correspond to $\omega_{2c}^+(q)$, $\omega_{2c}^-(q)$ and  $\omega_{2c2s}^-(q)$ in Ref.~\protect\citeSM{PhysRevB.85.165132SM} from top to bottom.   
  }
  \label{Skw-Nkw-T0}
\end{figure}

In this section, we revisit the dynamic structure factors (DSFs) of the half-filled Hubbard model for $U/\tH=8$ at zero temperature as a reference for comparison with the results in the main text for the pumped system. 
In general, the DSF at $T=0$ can be obtained via 
\begin{align}
S(q,\omega)=\sum_r e^{-\mathrm{i} qr} \int dt\,  e^{\mathrm{i}\omega t}
\langle \hat{O}_{j+r}(t) \hat{O}_{j}(0)\rangle\,.
\label{Sqw}
\end{align}
The most interesting point in terms of the $\eta$-pairing scheme is that the charge DSF $S_{\rm c}(q,\omega)$ with $\hat{O}_j=\hat{n}_j-1$ can be written using the $\eta$-operators in Eqs.~\eqref{etapm} and \eqref{etaz} as~\citeSM{PhysRevB.85.165132SM} 
\begin{align}
S_{\rm c}(q,\omega)&=4\sum_r e^{-\mathrm{i} qr} \int dt\,  e^{\mathrm{i}\omega t}
\langle \hat{\eta}^z_{j+r}(t) \hat{\eta}^z_{j}(0)\rangle
\\
&=2\sum_r e^{-\mathrm{i} (q+\pi)r} \int dt\,  e^{\mathrm{i}\omega t}
\langle \hat{\eta}^+_{j+r}(t) \hat{\eta}^-_{j}(0)\rangle\,, 
\label{Scqw}
\end{align}
where the momentum shift $\pi$ in Eq.~\eqref{Scqw} follows from the fact that the lattice translation operator anticommutes with $\eta^\pm$~\citeSM{EFGKK05SM}.

In Fig.~\ref{Skw-Nkw-T0} we demonstrate the numerical data obtained by the time-evolution technique with IBCs explained in Ref.~\citeSM{PVM12SM}, or `Method II' in Ref.~\citeSM{Zauner2015SM} instead of `Method I' used in the nonequilibrium case. 
The number of window sites is $L_{\rm W}=128$, and a Lorentzian broadening parameter $\eta_{\rm L}/\tH=0.1$ is used in the Fourier transformation in the time domain. 

Figure~\ref{Skw-Nkw-T0}(a) exhibits the spin DSF $S_{\rm s}(q,\omega)$ with $\hat{O}_j=\hat{S}_j^z$. 
As is well known, the spin excitations in the half-filled Hubbard chain are similar to those in the spin-1/2 Heisenberg antiferromagnet chain. The multispinon continuum lies between the lower [$\omega_{\rm low}(q)$] and upper [$\omega_{\rm up}(q)$] boundaries for two-spinon processes, which can be determined exactly by Bethe ansatz~\citeSM{EFGKK05SM}. Gapless spin excitations appear in $S_{\rm s}(q,\omega)$ at $q=0$ and $\pi$. 
In contrast, the charge DSF $S_{\rm c}(q,\omega)$ is gapped, reflecting the Mott-Hubbard gap $\Delta_{\rm c}$($\simeq4.68$ for $U/\tH=8$) as shown in Fig.~\ref{Skw-Nkw-T0}(b). Plotting $S_{\rm c}(q,\omega)$ on a log-scale, a two-holon continuum bounded by the dispersions $\omega_{\rm 2c}^\pm(q)$ from Bethe ansatz becomes visible~\citeSM{PhysRevB.85.165132SM}. 
The lower edge $\omega_{\rm 2c2s}^-(q)$, above which the spectral weight becomes nonzero, is defined by excitations with two holons and two spinons. 

The DSFs in the ramp-up regime of the pump field (Fig.~\ref{td-DSF} for $t\cdot\tH=5$) are similar to those without pumping but the latter are obtained with higher accuracy.

\section{Bond-dimension dependence of $A^-(\omega; t)$}
Because of the lack of exact analytical results for the pumped Mott insulator, it is difficult to assess the accuracy of our numerics at nonequilibrium. Here, we examine the bond-dimension ($\chi$) dependence of the integrated density of states $A^-(\omega;t)$ for the $\eta$-pairing state with $\omega_{\rm p}/\tH=7$ and $A_0=0.4$ at time $t\cdot \tH=15$, i.e., for the numerically most difficult situation considered in the main text. 

\begin{figure}[tbh]
  \includegraphics[width=0.45\linewidth]{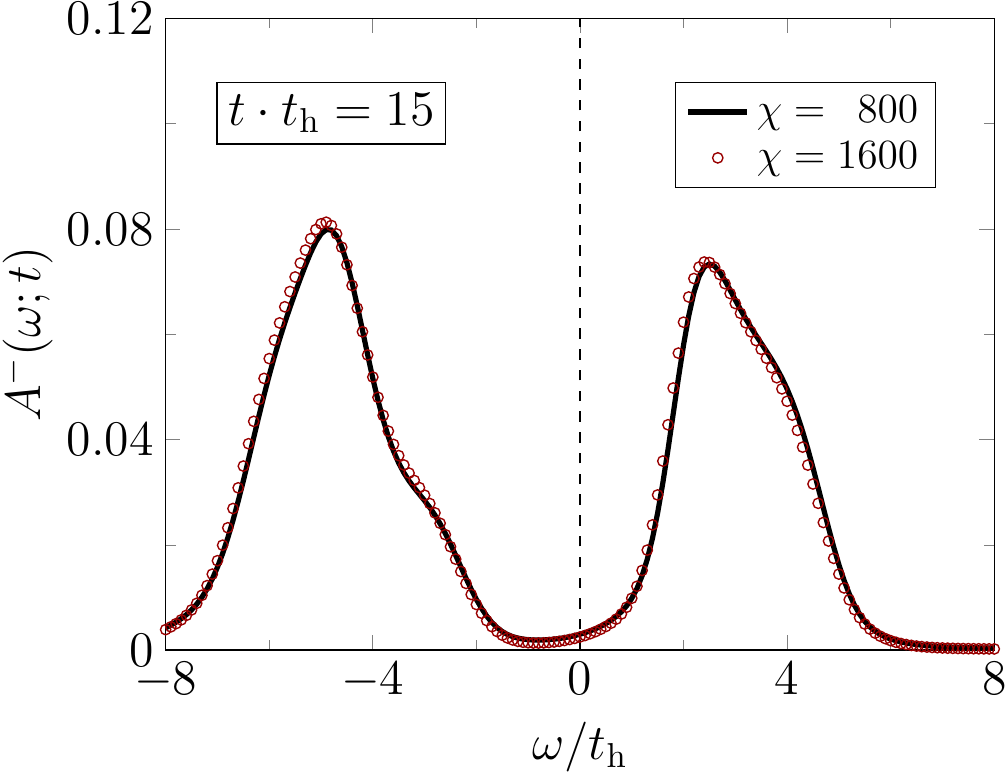}
  \caption{Bond-dimension dependence of the integrated density of states $A^-(\omega;t)$ with the pumping parameter set denoted as `+' in Fig.~\ref{contour-plots} at time $t\cdot\tH=15$. 
  }
  \label{Awt-t15}
\end{figure}

Figure~\ref{Awt-t15} shows $A^-(\omega; t)$ for bond dimensions $\chi=800$ and $1600$ for the above-mentioned parameter set. Obviously, the deviation of $A^-(\omega; t)$ is almost negligible and the shift of spectral weight from $\omega<E_{\rm F}$ to $\omega>E_{\rm F}$ is apparent in both cases.

%


\end{document}